\documentclass[prl,aps,twocolumn,superscriptaddress,showpacs,amsmath,amssymb,floatfix]{revtex4}

\usepackage[dvips]{graphicx}
\usepackage{bm}

\newcommand{\divr}{\bm\nabla\cdot}

\begin{document}

\title{Tkachenko Oscillations and the Compressibility of a Rotating Bose Gas}
\author{M.~Cozzini}
\affiliation{Dipartimento di Fisica, Universit\`a di Trento and BEC-INFM,
I-38050 Povo, Italy}
\author{L.~P.~Pitaevskii}
\affiliation{Dipartimento di Fisica, Universit\`a di Trento and BEC-INFM,
I-38050 Povo, Italy}
\affiliation{Kapitza Institute for Physical Problems, ul. Kosygina 2, 117334
Moscow, Russia}
\author{S.~Stringari}
\affiliation{Dipartimento di Fisica, Universit\`a di Trento and BEC-INFM,
I-38050 Povo, Italy}
  
\date{\today}

\begin{abstract}

The elastic oscillations of the vortex lattice of a cold Bose gas (Tkachenko
modes) are shown to play a crucial role in the saturation of the
compressibility sum rule, as a consequence of the hybridization with the
longitudinal degrees of freedom. The presence of the vortex lattice is
responsible for a $q^2$ behavior of the static structure factor at small
wavevectors $q$, which implies the absence of long range order in 2D
configurations at zero temperature. Sum rules are used to calculate the
Tkachenko frequency in the presence of harmonic trapping. Results are derived
in the Thomas-Fermi regime and compared with experiments as well as with
previous theoretical estimates.
 
\end{abstract}


\maketitle

In 1966 Tkachenko \cite{tkachenko} developed the theory of the elastic
oscillations of a vortical lattice in incompressible superfluids, predicting
the dispersion law $\omega_T=\sqrt{\hbar\Omega/4m}\,q$  where $m$ is the mass
of the fluid particles, $q$ is the wavevector of the wave and $\Omega$ is
the angular velocity of the fluid, related to the number $n_v$ of vortices per
unit surface by the relation $n_v=2\Omega m/h$.
The frequency $\omega_T$ is calculated in the reference frame rotating with
angular velocity $\Omega$. The Tkachenko modes are peculiar of superfluids,
where the formation of singly quantized vortices in the rotating liquid gives
rise to regular crystalline structures (Abrikosov lattice),
stationary in the rotating frame.
The theory of the Tkachenko modes was later developed by Baym \cite{baym},
who generalized the hydrodynamic theory of superfluids taking into account the
elasticity of the vortex lattice. The effects of the compressibility on the
Tkachenko modes were discussed in details by Sonin \cite{sonin}.
The compressibility changes dramatically the dispersion law at small
wavevectors $q$. In particular, when $q\ll\Omega/c$, where $c$ is the sound
velocity, the dispersion relation is no longer linear, but becomes quadratic in
$q$ as a result of the hybridization with the sound waves. 

The availability of vortex lattices in rapidly rotating Bose-Einstein
condensates \cite{ens mit,jilaHD} has stimulated new theoretical
studies \cite{anglin,baym03} of the Tkachenko oscillations in harmonically
trapped atomic gases.
In particular in Ref.~\cite{anglin} the discretized Tkachenko frequencies have
been calculated taking into account the finite and inhomogeneous nature of the
system.
In Ref.~\cite{baym03} the corresponding values of $q$ have been employed within
an improved dispersion law which includes the effects of compressibility. Fully
numerical calculations of the Tkachencko waves in harmonically trapped gases
have been also recently carried out using Gross-Pitaevskii theory
\cite{numerics}. The first experimental observation of the Tkachenko modes in
harmonically trapped gases has been recently reported by the group of JILA
\cite{jilaT}.
These modes represent the low energy counterpart of the hydrodynamic
modes already investigated experimentally in the presence of the
vortex lattice \cite{jilaHD}, in good agreement with theory \cite{cozzini}.

The main purpose of the present work is to show that, despite their elastic
nature, the Tkachenko oscillations can be naturally excited using density
perturbations, thereby opening new perspectives of experimental investigation. In
particular we will show that, in the homogeneous case, these modes
exhaust the compressibility sum rule in the limit of long wavelengths. In
harmonically trapped configurations the sum rule approach will be used to
calculate the frequency of the lowest azimuthally symmetric Tkachenko mode and
explicit results will be derived in the Thomas-Fermi regime. 

Let us start our discussion by evaluating the density response function of a
uniform gas containing a vortex lattice rotating in the $x$-$y$ plane with
angular velocity $\bm\Omega=\Omega\,\bm{e}_z$ at zero temperature, where
$\bm{e}_z$ is the unit vector along the $z$-direction.
In a compressible fluid the condition of uniformity can be fulfilled by
adding the harmonic potential term $m\Omega^2(x^2+y^2)/2$, which compensates 
the centrifugal effect produced by the rotation. This corresponds, in the
rotating frame, to using the Hamiltonian
\begin{equation} \label{eq:H}
H=\sum_{k=1}^N\,\frac{(\bm{p}-m\bm\Omega\wedge\bm{r})^2_k}{2m}+
\frac{1}{2}\,\sum_{i\neq j}g\,\delta(\bm{r}_i-\bm{r}_j) \ ,
\end{equation}
where $N$ is the number of atoms, $\bm{p}$ is the momentum and $g$ is the
coupling constant of the 2-body interaction.
The density response is easily evaluated using the coupled hydrodynamic-elastic
formalism~\cite{baym,sonin} characterized, in addition to the hydrodynamic
energy functional
\begin{equation} \label{eq:hd}
E_{\text{hd}} = \int\text{d}\bm{r}\,
\left[\frac{m}{2}\,(\bm{v}-\bm\Omega\wedge\bm{r})^2n+
\frac{1}{2}\,gn^2\right]
\end{equation}
where $\bm{v}$ is the velocity field in the laboratory frame and $n$ is the
density, by the elastic term
\begin{eqnarray}
\label{eq:elastic}
E_{\text{el}} & = & \int\text{d}\bm{r}\,\left\{2C_1(\divr\bm\epsilon)^2+\right. \\
\nonumber & & \left.+C_2\left[\left(\frac{\partial\epsilon_x}{\partial x}-
\frac{\partial\epsilon_y}{\partial y}\right)^2+
\left(\frac{\partial\epsilon_x}{\partial y}+
\frac{\partial\epsilon_y}{\partial x}\right)^2\right]\right\}
\end{eqnarray}
sensitive to the deformation of the lattice through the vortex displacement
field $\bm\epsilon$ and characterized by the elastic parameters $C_1$ and
$C_2$. At equilibrium one has $\bm{v}=\bm\Omega\wedge\bm{r}$ and
$\bm\epsilon=0$.
The dynamic response function takes the form
\begin{equation} \label{eq:chibaym}
\chi(q,\omega) = -N\,\frac{q^2\omega^2/m-\omega_+^2\omega_-^2/mc^2}
{[(\omega+i\eta)^2-\omega_+^2][(\omega+i\eta)^2-\omega_-^2]}
\end{equation}
with $\eta\to0^+$, where $\omega_+$, $\omega_-$ are, respectively, the upper
and the lower branches of the energy spectrum.
Expression~(\ref{eq:chibaym}) holds in the macroscopic regime
$q\ll(\hbar/m\Omega)^{-1/2}$ corresponding to wavelenghts larger than the
average distance between vortices.
The general expression for $\omega_{\pm}$ as a function of $C_1$ and $C_2$ has
been derived in Ref.~\cite{baym03}. Here we report the results in the
Thomas-Fermi regime $\hbar\Omega\ll mc^2$, corresponding to the condition that
the size $\hbar/mc$ of the vortex cores is much smaller than
$\sqrt{\hbar/m\Omega}$. In the Thomas-Fermi regime one has
$C_2=-C_1={n}\hbar\Omega/8$ \cite{baym03}. In this case the upper branch
follows the dispersion law $\omega^2_+=4\Omega^2+c^2q^2$ and exhibits a gap at
$q=0$. Conversely the low frequency branch, hereafter called Tkachenko branch
($\omega_-\equiv\omega_T$), obeys the gapless law \cite{sonin}
\begin{equation} \label{T}
\omega_T^2 = \frac{\hbar\Omega}{4m}\,\frac{c^2q^4}{4\Omega^2+c^2q^2} \ .
\end{equation}
For large $q$ Eq.~(\ref{T}) reproduces the original Tkachenko law
$\sqrt{\hbar\Omega/4m}\,q$, while for small $q$ it exhibits the quadratic
behavior $\omega_T=\sqrt{\hbar/16m\Omega}\,cq^2$.
The transition between the $q^2$ and $q$ dependence takes place at values
$q\sim\Omega/c$ which, in trapped condensates, can be significantly larger than
the inverse of the radial size of the system. This suggests that the effects of
compressibility, characterizing the $q^2$ dependence, play a crucial role in
the Tkachenko modes of a trapped gas, as explicitly pointed out in
Ref.~\cite{baym03}.

Starting from the density response function (\ref{eq:chibaym}) one can easily
calculate the energy weighted and the inverse energy weighted sum rules
relative to the density operator
$\rho_q=\sum_{k=1}^Ne^{-iqx_k}$.
These are given, respectively, by
$m_1(\rho_q)=\sum_n|\langle{n|\rho_q|0}\rangle|^2E_{n0}$ and
$m_{-1}(\rho_q)=\sum_n|\langle{n|\rho_q|0}\rangle|^2E_{n0}^{-1}$,
where $\sum_n$ is the sum over all the excited states and $E_{n0}=E_n-E_0$ is
the difference between the eigenenergies of the excited state $|n\rangle$ and
of the initial configuration $|0\rangle$ containing the vortex lattice.
The relation between these sum rules and the asymptotic behavior of the
density response function is given by \cite{PN}
$\chi(q,\omega)_{\omega\to\infty}=-2m_1(\rho_q)/(\hbar\omega)^2$ and
$\chi(q,0)=2m_{-1}(\rho_q)$.
In the first case one recovers the model independent $f$-sum rule
$m_1(\rho_q)=N\,(\hbar q)^2/2m$ which, at small $q$, is exhausted by the
high energy branch $\omega_+$. By taking the $\omega\to0$ limit of
$\chi(q,\omega)$ one instead finds the result $m_{-1}(\rho_q)=N/2mc^2$
also known as the compressibility sum rule.
The Tkachenko branch plays a crucial role in satisfying the latter sum rule. In
fact, because of the gap, the high energy branch $\omega_+$ contributes to
$m_{-1}(\rho_q)$ only through terms of order $q^2$. 

It is also worth noticing that the static structure factor
$S(q)=N^{-1}\sum_n|\langle{n|\rho_q|0}\rangle|^2$
is deeply affected by the rotation of the gas and
behaves like $q^2$, differently from what happens in non rotating interacting
fluids where it is linear in $q$ \cite{book}.
By using the zero temperature relationship
$NS(q)=(\hbar/\pi)\int_0^\infty\text{Im}\,\chi(\omega)\,\text{d}\omega$ one
finds the result
\begin{equation} \label{eq:Sq}
S(q) \,\ \to  \,\frac{\hbar q^2}{4m\Omega}\,
\left(1+\sqrt{\frac{\hbar \Omega}{4mc^2}}\right)
\end{equation}
when $q\to 0$, where the second term in the parenthesis is the contribution of
the Tkachenko branch
\cite{ft:S(q)}.
The corresponding suppression of the density fluctuations results in a dramatic
enhancement of the fluctuations of the phase of the order parameter, which
destroy long range order. This can be easily seen using the uncertainty
principle inequality $2S(q)(2n_q+1) \ge n_0$ \cite{PS91}, where $n_q$ is the
particle occupation number and $n_0$ is the Bose-Einstein condensate fraction.
Since $S(q) \to q^2$ one finds that $n_q$ diverges at least like $1/q^2$ at low
$q$, thereby ruling out Bose-Einstein condensation in 2D even at zero
temperature \cite{BEC}.

In the following we will use the sum rule technique to evaluate the Tkachenko
frequency through the ratio
\begin{equation} \label{eq:upper bound}
(\hbar\omega_T)^2 = \frac{m_1(F)}{m_{-1}(F)} =
\frac{\sum_n|\langle{n|F|0}\rangle|^2E_{n0}}
{\sum_n|\langle{n|F|0}\rangle|^2E_{n0}^{-1}}
\end{equation}
between the energy weighted and inverse energy weighted sum rules relative to
the excitation operator $F$. Our final goal is to derive explicit results in
the presence of harmonic trapping. Eq.~(\ref{eq:upper bound}) provides a
rigorous upper bound to the frequency of the lowest energy mode excited by $F$.
The proper choice of the operator is a crucial step in the calculation. For
example, in the uniform case it would not be appropriate to use the density
operator $\rho_q$ for $F$ since, as already pointed out, the $f$-sum
rule $m_1(\rho_q)$ is exhausted by the upper branch at small $q$ and the ratio
(\ref{eq:upper bound}) would not coincide with the Tkachenko frequency. 
 
The general strategy for the identification of the excitation operator is
suggested by the fact that the Tkachenko modes have zero energy in the
hydrodynamic approximation \cite{chevy}, where elasticity effects are ignored.
So one should look for excitation operators $F$ whose energy weighted sum rule
$m_1(F)\!=\!\langle{[F^{\dagger},[H,F]]}\rangle/2$
vanishes when evaluated in the hydrodynamic approximation.

In a uniform system this condition is satisfied by the non local choice
$F=\rho_q-i(q/2m\Omega)\sum_{k=1}^N[e^{-iqx}(p_{y}-m\Omega x)]_k$.
In fact the corresponding double commutator takes the form
$[F^{\dagger},[H,F]]=(\hbar^2q^4/4m^3\Omega^2)\sum_{k=1}^N(p_{y}-m\Omega x)_k^2$
and its expectation value identically vanishes if one uses the hydrodynamic
prescription $\bm{p}=m\bm{v}(\bm{r})$ with the equilibrium condition
$\bm{v}=\bm\Omega\wedge\bm{r}$.
The elastic contribution to $m_1(F)$ can be conveniently calculated applying to
the equilibrium configuration the unitary transformation $U=e^{i\theta S}$,
where $S=(F+F^{\dagger})/2$ and $\theta$ is a small parameter. Due to the
presence of the non local transverse term $e^{-iqx}p_{y}$ in $F$, the
transformation $U$ gives rise to the vortex displacement
$\bm\epsilon(\theta)=-\theta(\hbar q/2m\Omega)\sin(qx)\bm{e}_y$, where
$\bm{e}_y$ is the unit vector along the $y$-direction. By calculating the
corresponding elastic energy change (\ref{eq:elastic}) and considering the
expansion $E(\theta) = \langle{U^{-1}HU}\rangle \simeq
E_0+\theta^2\langle{[S,[H,S]]}\rangle/2$ for the total energy of the
system, one derives the result
$m_1(F)=N(\hbar\Omega/4m)(\hbar^2q^4/8\Omega^2m)$.

To obtain $m_{-1}(F)$ we calculate the static response $\chi(F)$ to the
perturbation $-\lambda F+\text{h.c.}$ using the hydrodynamic energy functional
and the relationship $\chi(F)=2m_{-1}(F)$. The elastic term provides higher
order corrections. For small values of $q$ the leading contribution arises from
the density component $\rho_q$ in the Tkachenko operator and is fixed by the
compressibility of the gas. For larger values of $q$, or for incompressible
fluids, the static response is instead determined by the transverse current
term $e^{-iqx}(p_{y}-m\Omega x)$. In general one finds the result
$m_{-1}(F)=(N/8\Omega^2mc^2)(4\Omega^2+c^2q^2)$. Using the above results for
$m_1(F)$ and $m_{-1}(F)$ Eq.~(\ref{eq:upper bound}) reproduces exactly the
Tkachenko dispersion law~(\ref{T}).

Let us now discuss the case of a harmonically trapped rotating gas, where
the non interacting part of the Hamiltonian (\ref{eq:H}) should be replaced by
$\sum_{k=1}^N[p^2/2m+
m(\omega_\perp^2r_\perp^2+\omega_z^2z^2)/2-\bm\Omega\cdot(\bm{r}\wedge\bm{p})]_k$
\cite{ft:HDfunc in trap}
with $r_{\perp}=\sqrt{x^2+y^2}$.
A first estimate of the compressional effects on the frequency of the lowest
Tkachenko mode was obtained \cite{baym03} using the dispersion
law~(\ref{T}) and setting $q=\alpha/R_\perp$, where $R_\perp$ is the
Thomas-Fermi radial size of the gas
\cite{ft:3Dradius}
and $\alpha$ is a dimensionless parameter
characterizing the discretization of the normal modes.
The value $\alpha=5.45$ for the lowest azimuthally symmetric mode was extracted
from the result of Ref.~\cite{anglin} holding in the incompressible regime
where the frequency is linear in $q$.
Since in a trapped gas the relevant excitations are affected by the
compressibility of the gas, it is not obvious that the above estimate
is enough accurate and it is hence important to have more precise calculation
of the frequency of the lowest Tkachenko mode, taking into account the finite
size, the inhomogeneity as well as the compressibility of the gas. In the
following we will provide such an estimate using the method of sum rules. We
will make the ansatz $F=\sum_{k=1}^N[P(r_\perp)\ell_z+Q(r_\perp)]_k$ for the
operator exciting the lowest Tkachenko mode. Here $\ell_z$ is the $z$-direction
component of the angular momentum operator and $P$ and $Q$ are real functions.
Following the same considerations made in the uniform case, we find that the
average value of the double commutator
$[F^{\dagger},[H,F]]=(\hbar^2/m)\sum_{k=1}^N(P'\ell_z+Q')_k^2$ exactly vanishes
in the hydrodynamic approximation, where $\ell_z=m\bm{r}\wedge\bm{v}$ with
$\bm{v}=\bm\Omega\wedge\bm{r}$ at equilibrium, provided the condition
\begin{equation} \label{eq:Q'}
P'm\Omega{}r_\perp^2+Q'=0 
\end{equation}
is satisfied. As in the uniform case, one can then introduce the unitary
transformation $U=e^{i\theta F}$, which produces the vortex displacement
$\bm\epsilon(\theta)=\hbar\theta P\bm{e}_z\wedge\bm{r}$. Expanding
$E(\theta)=\langle{U^{-1}HU}\rangle$ in $\theta$, the $m_1(F)$ sum rule is
given by the second order term of the expansion which can be calculated using
the elastic energy change~(\ref{eq:elastic}). One finds the result
$m_1(F)=\hbar^2(\hbar\Omega/8){\langle{{P'}^2r_\perp^2}\rangle}$.
The $m_{-1}(F)$ sum rule is instead extracted from the static hydrodynamic
response. The variations of the density and of the velocity induced by the
static perturbation $-\lambda F$ take the form
\cite{ft:chevy}
\begin{eqnarray}
\label{eq:dn}
\delta{n} & = & [\delta\Omega\,m\Omega{}r_\perp^2+\delta\mu+
\lambda(Pm\Omega{}r_\perp^2+Q)]/g \ , \\
\label{eq:dv}
\delta\bm{v} & = & (\delta\Omega+\lambda P)\,\bm{e}_z\wedge\bm{r}\ ,
\end{eqnarray}
where the changes $\delta\Omega$ and $\delta\mu$ are obtained imposing the
conservation of the number of particles and of angular momentum. After some
straightforward algebra one finds:
$m_{-1}(F) = m\left(\langle{P^2r^2_\perp}\rangle-
{\langle{Pr_\perp^2}\rangle}^2/\langle{r^2_\perp}\rangle\right)/2+
\int\text{d}\bm{r}\,(Pm\Omega r_\perp^2+Q)\,\delta{n}/2\lambda$.

\begin{figure}
\includegraphics[width=8.5cm]{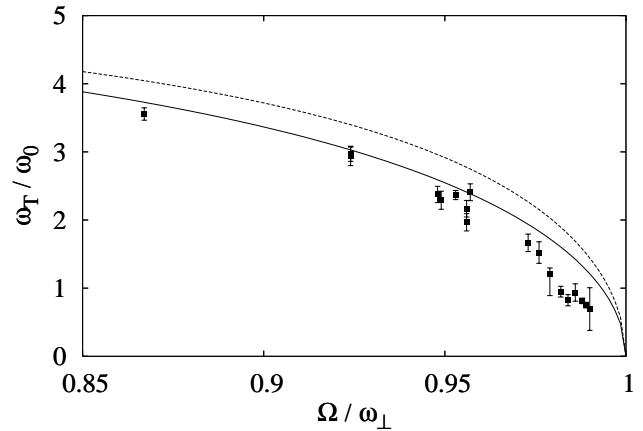}
\caption{\label{fig:1}Lowest Tkachenko frequency of a trapped Bose gas in units
of $\omega_0=\sqrt{\hbar\Omega/4mR_\perp^2}$ as a function of the angular
velocity $\Omega$.
The full line is the sum rule result, while the dashed line is the
prediction of Eq.~(\ref{T}) with $q=5.45/R_\perp$ (see text). Experimental
points are taken from Refs.~\cite{jilaT,jilaLLL} using the 3D Thomas-Fermi
value for $R_\perp$. The prediction of Ref.~\cite{anglin} corresponds to
$\omega_T/\omega_0=5.45$.}
\end{figure}

The above results hold for any choice of the functions $P$ and $Q$ satisfying
the condition (\ref{eq:Q'}). We have written $P$ as a polynomial expression of
the form $P=\sum_sp_s(r_\perp/R_\perp)^s$ and analogously for $Q$.
Using the Thomas-Fermi equilibrium profile
$n=\{\mu-[(\omega_\perp^2-\Omega^2)r_\perp^2+\omega_z^2z^2]/2\}/g$,
where $\mu=m(\omega_\perp^2-\Omega^2)R_\perp^2/2$ is the chemical
potential, all the integrals involved in $m_1(F)$ and $m_{-1}(F)$ are
analytical and one can write the frequency in the form
\begin{equation} \label{eq:omega^2}
\omega^2_T = \frac{\hbar\Omega}{4m}\,\frac{1}{R_\perp^2}\,
f(\Omega/\omega_\perp) \ .
\end{equation}
We have determined the frequency of the lowest mode by minimizing the function
$f$ with respect to the coefficients $p_s$ at fixed $\Omega/\omega_\perp$. In
practice good convergence to the exact value of $\omega_T$ is already ensured
for $s=3$. The results are reported in Fig.~\ref{fig:1}.
When $\Omega\ll\omega_\perp$ the function $f$ approaches a constant value. When
$\Omega\to\omega_\perp$ it instead vanishes like $1-\Omega^2/\omega_{\perp}^2$.
Notice, however, that if $\Omega$ is too close to $\omega_\perp$ the validity
of the present calculation, based on the Thomas-Fermi approximation, breaks
down due to the diluteness of the gas produced by the rotation. In
Fig.~\ref{fig:1} we also show the prediction obtained using Eq.~(\ref{T}) with
$q=\alpha/R_\perp$ and $\alpha=5.45$. Using the relation $mc^2=\mu$ it is
immediate to verify that also in this case the dispersion law can be factorized
in the form~(\ref{eq:omega^2}) with $f(x)=(1-x^2)/[a^{-1}(1-x^2)+b^{-1}x^2]$
and $a=\alpha^2=29.7$, $b=\alpha^4/8=110.3$. The results of the sum rule
approach instead correspond to $a=31.3$ and $b=75.8$
\cite{ft:2D}.
The relative difference between the two predictions is more and more pronounced
as $\Omega\to\omega_\perp$ pointing out the inadequacy of the choice
$q=5.45/R_\perp$ when $R_\perp$ is large and $q$ becomes small.
Fig.~\ref{fig:1} also shows that the sum rule prediction is systematically
closer to the experimental values and that the deviations of the measured
Tkachenko frequencies from the Thomas-Fermi values become important only for
values of $\Omega$ very close to $\omega_\perp$, where quantum Hall effects
should be taken into account \cite{BEC,baym03,jilaLLL}.

\begin{figure}
\includegraphics[width=8.5cm]{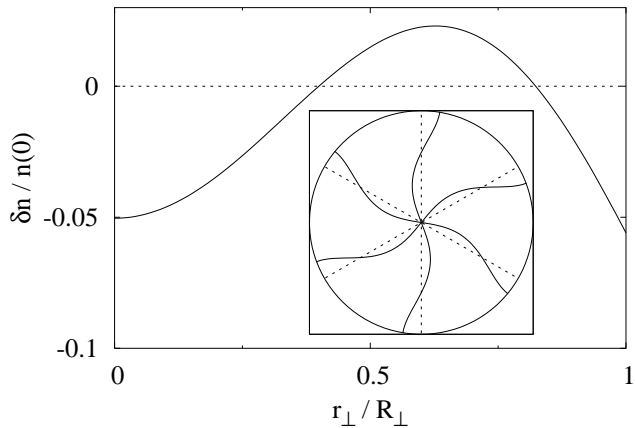}
\caption{\label{fig:2}Density change of the lowest Tkachenko mode at
$\Omega/\omega_\perp=0.9$. The inset shows the corresponding displacement
$\epsilon$ of the vortex lattice at the angles $\pi/6,\pi/2,5\pi/6$. The
amplitude of the oscillation corresponds to $\delta\Omega/\Omega=0.05$.}
\end{figure}

In Fig.~\ref{fig:2} we report the shape of the density
deformation~(\ref{eq:dn}) associated with the Tkachenko oscillation. A density
perturbation of this form, produced by a proper change of the trapping
potential, should result in a significant excitation of the Tkachenko mode.
This density change differs from the scaling deformation
associated with the radial breathing hydrodynamic mode.
Actually the breathing and the Tkachenko modes are orthogonal as confirmed by
the fact that the density variation~(\ref{eq:dn}) does not result in any change
of the average square radius: $\int\text{d}\bm{r}\,r_\perp^2\,\delta{n}=0$. In
Fig.~\ref{fig:2} (inset) we also show the amplitude of the corresponding
vortex lattice deformation, obtained from the relation
$\bm\epsilon=\delta\bm{v}/\omega_T$. Similar shapes have been obtained in the
theoretical calculations of Refs.~\cite{anglin,numerics} and observed
experimentally in Refs.~\cite{jilaT,jilaLLL}.

In conclusion we have shown the occurrence of important compressional features
exhibited by the Tkachenko oscillations in rotating Bose gases. Sum rules have
permitted to provide accurate estimates of the frequency of the lowest mode
allowing for a detailed comparison with experiments.

Useful discussions with Gordon Baym, Jean Dalibard and Peter Engels are
acknowledged.


\begin{thebibliography}{99}
\bibitem{ft:S(q)}
Result~(\ref{eq:Sq}) holds in the $\hbar\Omega \ll mc^2$ limit. In
the general case one has $S(q)\to(\hbar q^2/4m\Omega)(1+\sqrt{2C_2/nmc^2})$.
\bibitem{ft:HDfunc in trap}
The first term in the hydrodynamic functional~(\ref{eq:hd}) should be
correspondingly replaced by $\int\text{d}\bm{r}\,n
[m{\bm v}^2/2+m(\omega_{\perp}^2r^2_{\perp}+
\omega_z^2z^2)/2-\bm\Omega\cdot(\bm{r}\wedge m\bm{v})]$.
\bibitem{ft:3Dradius}
In 3D one has $R_\perp=a_\perp
[15Na\omega_z/a_\perp\omega_\perp(1-\Omega^2/\omega_\perp^2)]^{1/5}$, where $a$
is the scattering length and $a_\perp=\sqrt{\hbar/m\omega_\perp}$.
\bibitem{ft:chevy}
Notice that Eq.~(\ref{eq:Q'}) can also be obtained from
Eqs.~(\ref{eq:dn}-\ref{eq:dv}) imposing the condition $\delta\bm{v} =
(\bm\Omega{}/2\Omega{}^2)\wedge\bm\nabla(g\,\delta{n}/m)$, which characterizes
a class of zero energy hydrodynamic solutions. This points out explicitly
the relation between these special hydrodynamic solutions and the Tkachenko
modes~\cite{chevy}.
\bibitem{ft:2D}
These value are obtained evaluating the integrals with 3D
Thomas-Fermi density profiles. If one uses 2D Thomas-Fermi profiles one finds
$a_{2D}=29.5$ and $b_{2D}=92.8$. Notice that the $a_{2D}$ is very close to the
value $a=\alpha^2=29.7$ extracted from the 2D calculation of
Ref.~\cite{anglin}.
\end{thebibliography}
\end{document}